# Biologically Inspired Nanomaterials – A Conference Report

Melik Demirel[1,*], Atul Parikh[2], Vincent Crespi[3], Scott Reed[4,*]

1. College of Engineering, The Pennsylvania State University, University Park, PA, 16802
2. College of Engineering, University of California, Davis, CA, 95616
3. College of Science, The Pennsylvania State University, University Park, PA, 16802
4. College of Science, Portland State University, Portland, OR, 97207

**ABSTRACT**

The understanding of the nanoscale physical properties of biomolecules and biomaterials will ultimately promote the research in the biological sciences. In this review, we focused on theory, simulation, and experiments involving nanoscale materials inspired by biological systems. Specifically, self-assembly in living and synthetic materials, bio-functionalized nanomaterials and probing techniques that use nanomaterials are discussed.

*Corresponding authors: E-mail: mdemirel@engr.psu.edu (MCD), sreed@pdx.edu (SR)

How can inspiration from nature improve the design of synthetic materials? Direct observation of nanoscale materials has recently become commonplace and it is already evident that the observation of biological nanostructures is having an influence on the design of synthetic materials. Some of the successful nanoscale structures either use or mimic biological materials in their designs. [1] An emerging frontier is this interface between biology and nanomaterials. It is with this common interest that an interdisciplinary group of scientists convened in November of 2005 at the first ICAM workshop on "Biologically Inspired Nanomaterial." The workshop focused on understanding the nanoscale physical properties of biomolecules and biomaterials that will ultimately aid research in the biological sciences. This report is a brief summary of the research topics that are presented at the ICAM workshop.

We can create polymers that behave like antibodies, [2] surfaces that possess the surface properties of cells, [3, 4] and synthetic motors that can be harnessed for local power [5] by mimicking successful elements of natural nanoscale architectures. Designing environmentally benign systems is also essential to minimizing the impact of future technologies. Here too, natural systems can serve as an inspiration for their ability to operate using renewable resources.

How we better mimic nature's solutions in designing the electronic, optical, and architectural components of materials? Similar questions have been asked since the invention of the optical microscope. Although we are not the first to ask these questions, this is an appropriate time to re-visit these questions. We have begun to view the world at nanometer length scales, and novel theoretical models are being developed to describe nature at this level. The invention of the light microscope forever changed our view of the natural world. Although it was designed as a tool of observation, since its invention five centuries ago the technological landscape has altered dramatically as a result. The images of cells, diatoms, bacteria, *etc* influenced artists, architects, engineers, and scientists. With novel tools and models available to us, it is an appropriate time to explore both how these will alter our understanding of the natural world that is highly complex, (see Figure 1) and also how the materials we synthesize will be influenced by the new views.

Allara's group has reported on the use of self-assembly to construct well organized materials relying on noncovalent and covalent interactions between relatively small precursor molecules. [1,

[6] This "bottom-up" approach to supramolecular organization leads to unusual and important properties in biologically derived materials (e.g. nanoparticles and nanowires). Although these processes are not well understood, they seem to be critical in the biological assembly of complex material systems.

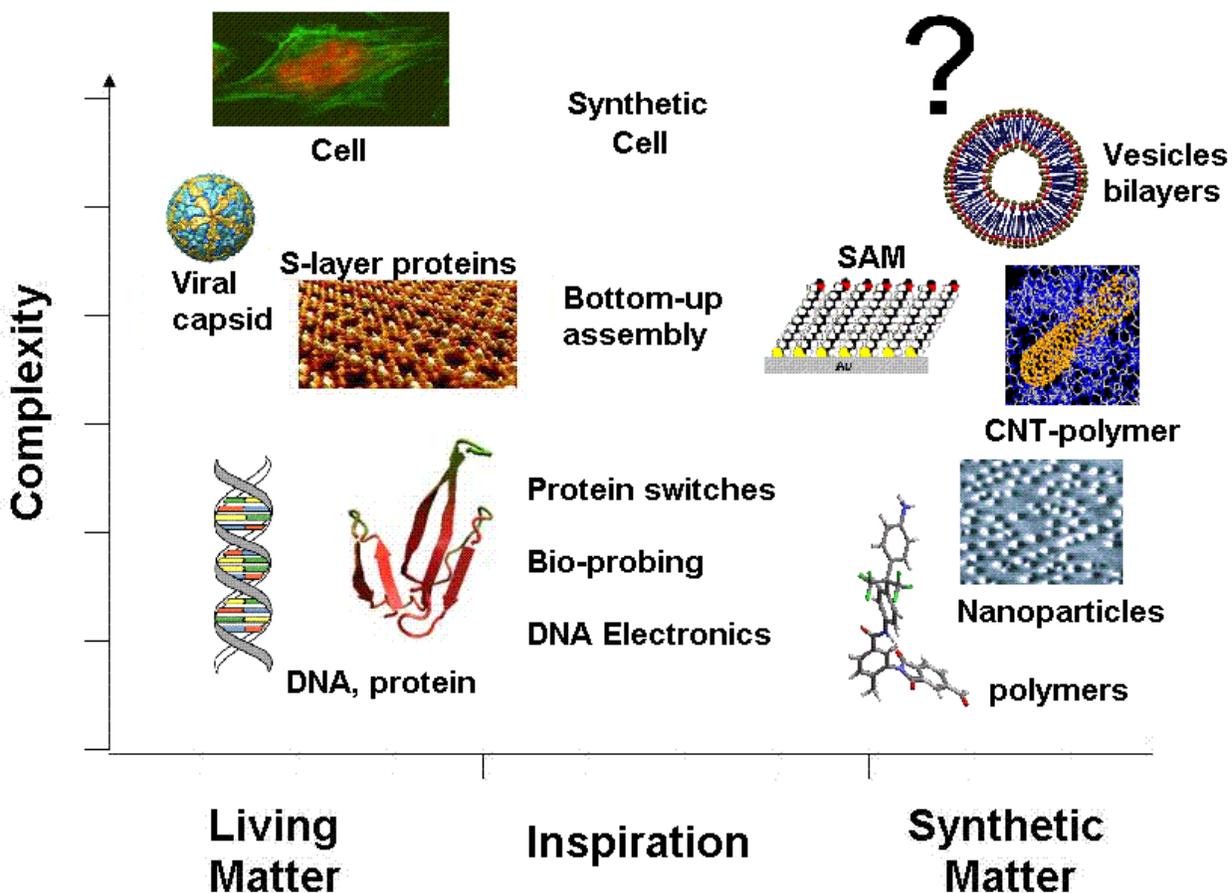

**Figure 1**. Synthetic matter is inspired from living matter at different scales (i.e. nanometer to micrometer) and complexities.

Nature has a well diversified counterpart in synthetic self-assembly. Using molecular machinery, living cells build complex molecules and organelles. For example, Sleyter and coworkers have showed that S-layer proteins can be assembled as building blocks similar to the bottom-up self assembly approaches by creating periodic crystalline proteins (from *B. sphaericus*), with center-to-center spacing of 13.1 nm. [7, 8] The building blocks of S-layer protein promises a future for molecular electronic, optical, and magnetic devices. [9]

Functionalized nanoparticles are being developed for many applications ranging from medicine to electronics [2, 10-13]. Willner and coworkers have demonstrated an approach to electrically contact redox enzymes with their macroscopic environment by reconstitution of an apoezynme with an FAD-gold nanoparticle conjugate. The ability to control the shape and structure of redox enzymes (such as apo-glucose oxidase) make biomolecules attractive building blocks for functional circuitry devices. [14] Hutchison and coworkers developed methods of nanofabrication based upon the assembly of functionalized nanoparticles. One method, biomolecular nanolithography, involves self-assembly of nanoparticles onto biopolymeric (DNA) scaffolds to form lines and more complex patterns. [15] Using DNA as a model, Williams and coworkers [16] synthesized artificial oligomeric peptides that form duplexes upon coordination of metal atoms that are designed to mimic the conformational and functional properties of DNA.

The thermodynamic and conformational properties of proteins have been studied for several decades but the vast majority of such studies were done using proteins in solution. The anisotropic and inhomogenous physical properties of proteins have been addressed by Scoles' group [17, 18] recently. They have studied these questions through the specific surface confinement of one or more patches of proteins, each containing about 100-1000 elements. They have used alkanethiol SAMs, $CH_3-(CH_2)_{17}-SH$, on gold surfaces and have created nanopatterns using the AFM based nanografting technique and produced patches of the *de-novo* S-824 protein. [17]

Banavar and coworkers addressed an essential difference between synthetic and living matter from a theoretical perspective. Both kinds of matter are governed by physical law. [19] While the gross behavior of synthetic material is reasonable understood, a similar simple understanding, even in principle, has been missing for living matter. Recent development of suitable coarse-grained (often topology-based) models captures the main features of the kinetics, thermodynamics, and the elastic response of various biomolecules (e.g. protein, DNA). [20, 21] For example, analysis of X-ray crystal structures has clarified the nature of antibody-antigen interactions, and the conformational basis of specificity and affinity, but does not provide a clear picture of the dynamics of antigen recognition. Demirel and Lesk [21] showed that the unligated state of the primary antibody has a well-defined structure, fluctuating no more widely than that of the secondary antibody, and undergoes a discrete structural rearrangement in response to ligand.

A detailed understanding of proteins may establish building blocks for future nano-scale diagnostic or therapeutic devices. Hancock's laboratory [5] is concentrating on the kinesin superfamily of microtubule-based motors, which are involved in a broad array of cellular processes including axonal transport, the positioning of intracellular organelles, and the movement of chromosomes during cell division. [22]

Self-assembly also exists at the microscale, for example cellular membranes. Model membrane system (e.g. bilayers) and their dynamics have been studied extensively by several groups [23, 24]. The Boxer group has addressed the formation of domains and protein association with these domains. This work has motivated the development of advanced optical microscopy methods for probing the interface between membranes on solid supports and cell membranes, potentially with nanometer vertical resolution. [23, 25] The Weiss group is able to induce separation of components in membranes by manipulating their environment and applying forces to them. [26] They have worked on model systems such as giant unilamellar phospholipid bilayer vesicles as well as on true biological systems. They relate membrane curvature to the local composition of multi-component lipid bilayers. Such variations within membranes are important biologically for such processes as including exocytosis, endocytosis, and adhesion. [26]

Current approaches to control cell behavior through micro- and nano-engineered materials have applications including medical implants, cell supports, and materials that can be used as three-dimensional environments for tissue regeneration. Austin and coworkers have recently measured the dynamic traction forces exerted by epithelial cells on a microengineered substrate. [27] Traction forces induced by cell migration are deduced from the measurement of the bending of these pillars and are correlated with actin localization by fluorescence microscopy. Their data provide definite information on mechanical forces exerted by a cellular assembly. Maximal-traction stresses at the edge of a monolayer correspond to higher values than those measured for a single cell and may be due to a collective behavior. Demirel and coworkers recently deposited a new class of bioactive nanoengineered sculptured thin films [28] whose morphology as well as material properties have been demonstrated to overcome some of the incompatibility problems. Goldstein's group has focused on the self organization at a cellular level. Using the geometry of a sessile drop they have demonstrated in suspensions of *B. subtilis* the self-organized generation of a persistent hydrodynamic vortex which traps cells near the contact line. [29]

In conclusion, the use of biological principles is becoming widespread in the design of nanomaterials. It is clear that the knowledge of nanoscale physical properties of biomolecules and biomaterials is essential to progress in biology, and will engender progress in physics.

## ACKNOWLEDGEMENT

The biologically inspired Nanomaterials workshop is supported by funds from the Institute for Complex Adaptive Matter, the Materials Research Institute of the Pennsylvania State University, and the Office of Sponsored Projects and Research at Portland State University. We thank to Dr. David Allara and Dr. Akhlesh Lakhtakia for critical review of the material presented in this review and Dr. David Pines for many constructive discussions.